\documentclass[12pt]{iopart}
\usepackage{iopams}
\usepackage{graphicx}
\begin{document}

\title{The Aladin2 experiment: status and perspectives}

\author{Giuseppe Bimonte\dag, Detlef Born\dag, Enrico Calloni\dag 
\footnote{Talk given by this author.},  Giampiero Esposito\dag, 
Uve H\"{u}bner\ddag,
Evgeni Il'ichev\ddag, Luigi Rosa\dag, Ornella Scaldaferri\dag, 
Francesco Tafuri\dag, Ruggero Vaglio\dag}

\address{\dag INFN, Sezione di Napoli, and Dipartimento di 
Scienze Fisiche, Complesso Universitario di Monte
S. Angelo, Via Cintia, Edificio N', 80126 Napoli, Italy}

\address{\ddag Institut F$\ddot{u}$r Physikalische HochTechnologie e. V., 
Postfach 10 02 39 - 07702 Jena, Germany}

\begin{abstract}  
Aladin2 is an experiment devoted to the first measurement of variations 
of Casimir energy in a rigid cavity. The main scientific 
motivation relies on the possibility of the first 
demonstration of a phase transition influenced by vacuum 
fluctuations. The principle of the measurement, based on the 
behaviour of the critical field for an in-cavity superconducting 
film, will be only briefly recalled, being discussed in detail 
in a different paper of the same conference (G. Bimonte et 
al.). In this paper, after an introduction to the long term 
motivations, the experimental apparatus and the results of the 
first measurement of sensitivity will be presented in 
detail, particularly in comparison with the expected signal. 
Last, the most important steps towards the final measurement will be discussed. \end{abstract}

\section{Introduction}

The last decade has seen impressive improvements in the 
measurements of Casimir force
\cite{Carugno} and this has triggered renewed interest 
in more general direct measurements of 
vacuum fluctuation effects.
In a recent paper \cite{Callo}, pointing out the lack 
of any experimental verification of the vacuum energy 
gravitational interaction, we noticed that the present macroscopic small 
force detectors, like the gravitational wave interferometers, 
might have the sensitivity to measure the extremely small forces 
exerted by the earth gravitational field on a suitable Casimir cavity. 
As we pointed out, the possibility of success is linked both to the 
realization of a  many-cavities layered rigid structure and to an 
efficient modulation of the Casimir energy. As an example, for a 
$10^6$-cavities structure consisting of alternate layers of aluminum 
(100 nm) and alumina (5 nm), with 0.5 modulation depth and tens Hz 
modulation frequency, the signal might be detected in one month 
integration time with a signal-to-noise ratio of about 100. The 
peculiar properties of such cavities, jointly with modulation depths, 
make the measurement virtually impossible at this moment of time; 
nevertheless, the compatibility of such experimental conditions with a not 
too optimistic progress in film depositions has ``triggered'' us in 
searching for methods for the modulation of Casimir energy in a rigid 
cavity without, of course, exchanging with the system an energy too 
bigger than the modulated Casimir energy itself, to avoid destroying 
any possibility of measurement and control. In this spirit we have analyzed 
the possibility to induce variations of energy by realizing the cavity 
mirrors with materials that can undergo superconducting transitions. A variation of Casimir energy is expected because the mirrors' 
reflectivity changes, while other exchanged energy is expected to be small, 
being linked to the condensation energy. The use of a phase transition 
offers not only the possibility to perform the energy variation, 
but also an interesting method to measure it. If the condensation 
energy and the variation of Casimir energy are of comparable magnitude it 
can be expected that the latter may have a measurable effect on the 
transition itself. This is indeed the case if the transition is obtained 
by means of an external applied magnetic field. For a given temperature, 
the external field needed to destroy superconductivity, i.e. the 
critical field, is in fact proportional  to the total variation in free 
energy between the normal and superconducting state at zero field: if 
the condensation energy and Casimir variation are comparable, the 
total energy variation, and thus the critical field, of a film being part of 
a cavity can be sensibly different from that of a simple film. The 
Aladin2 experiment has been conceived to verify this hypothesis, 
demonstrating the effect of vacuum fluctuations on a phase transition; the 
study of the possibility to modulate Casimir energy to verify its 
gravitational interaction, which was the original starting point, remains 
as a long term motivation. The project has been funded by the Italian 
INFN (Istituto Nazionale di Fisica Nucleare) and recently has been 
joined by the German IPHT (Institut F$\ddot{u}$r 
Physikalische HochTechnologie). The 
final measurement is foreseen for the end of 2007. Although the ideal 
cavity would be a five-layer structure \cite{Moste, Bimon1}, 
the actual cavity is a three-layer structure that warrants a safer 
realization and electrical contacting: a thin film of superconducting 
metal, a dielectric layer and a final film of normal metal. The cavity is 
placed at cryogenic temperature and an external magnetic field is 
applied, parallel to the plane of the films. The applied field necessary 
to destroy superconductivity $H^{C}_{||}(T)$ is measured as a function 
of temperature. The expected signal is a different behaviour of the 
function $H^{C}_{||}(T)$ with respect to the critical 
field $H^{F}_{||}(T)$ of simple film. Details on calculations and on 
other theoretical aspects can be found in a proceedings paper of this 
same conference \cite{Bimon1}. In the present paper, after recalling 
the expected signal for the actual experimental configurations, we will 
report on our present sensitivity and next experimental steps. 

\section{Expected signal and sensitivity limits}

As shown in detail in \cite{Bimon1}, a good choice for the cavity 
configuration is a  structure having a first superconducting 
film of 5 nm thickness, a second dielectric layer 10 nm thick and a final 
metal layer 100 nm thick. The shift in magnetic field is maximized for 
low condensation energies, so that superconductors having a low $T_{c}$ 
should be preferable. Nevertheless, measurements at very low temperatures
could be particularly difficult and time consuming, not easily allowing 
a high number of measurements and statistical analysis. As a good 
compromise between amplitude of expected signal and multiple measurements 
we choose to work in the 1 K region of temperatures, where the cooling down 
time is relatively short and the measurements can be performed on typically 
well known soft superconductors (like Aluminum, Zinc, Tin or even 
Beryllium if deposited on cooled substrate \cite{Adams}). 
To illustrate the expected 
signal, in fig. 1 $H_{||}(T)$ is reported for a single film (dashed curve) and 
for an in-cavity film (green and black curves) in the optimal 
configuration of first 
and third layers made of Beryllium and the intermediate layer of a 
native oxide. 

\begin{figure}
	\centering
		\includegraphics[width= 8 cm]{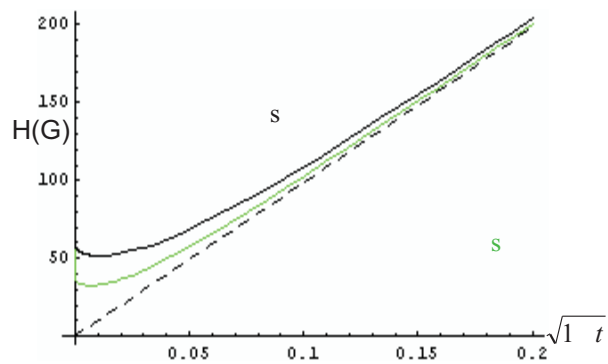}
	\caption{Critical field of Beryllium; simple and in-cavity film}
	\label{fig:berillio}
\end{figure}

As is seen in the figure the in-cavity film, for 
reduced temperature $t = \frac{T}{T_{c}}$ approaching unity (not valid in 
a neighbourhood \cite{Bimon2}), should exhibit an $H^{C}(t)$ which deviates 
from the usual law $H^{F}_{||}(t) \propto \sqrt{1 - t}$ valid for single 
film. The ratio of the field shift and simple 
film field $r = \frac{H^{F}(t) - H^{C}(t)}{H^{C}(t)}$ is five-tens percent 
for reduced temperature sufficiently far from $T_{c}$, so that the 
signal should be quite easily measurable. Unfortunately, although Beryllium 
is a very promising material, its toxicity makes it difficult to find it in 
the market (at least properly deposited) and also to be home deposited. For 
this reason we have decided to start the experimental work with more 
handily materials like Aluminum and Zinc. In particular, Aluminum has been 
chosen since it is a very well known material and can be used to test the 
sensitivity of the experimental apparatus, while Zinc is expected to have a 
sufficiently low condensation energy to exhibit a measurable signal. 
Calculations show that the nature of the intermediate oxide layer does not 
affect sensibly the field shift, while metallic properties of the third 
layer, in particular plasma frequency and mean free path, heavily 
do \cite{Bimon1, Bimon2}.\\

The situation is described in fig. 2, where the ratio $r$ is reported for 
various configurations:  first layer of Aluminum and third layer of 
Gold (red curve), Zinc and Gold (black curve),  Zinc and 
``perfect reflector mirror'' (blue curve).  
  
\begin{figure}
	\centering
		\includegraphics[width= 8 cm]{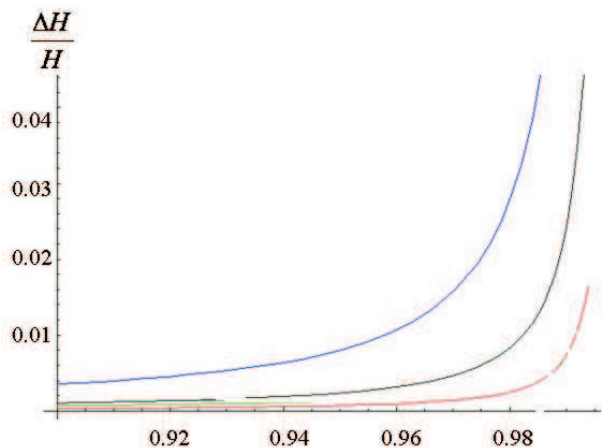}
	\caption{Expected relative field deviation }
	\label{fig:teorica}
\end{figure}

The work of our group is also devoted, at present, to discovering which 
materials have a sufficiently high plasma frequency and mean free path 
to approach the behaviour of a perfect mirror. Among the metals Beryllium 
is the best but, again in light of its toxicity, further analysis is 
devoted to finding whether some alloy or compound might be used instead. From 
fig. 2 it can be seen that, in case of Zinc with not optimal mirror 
reflectivity (Gold), the sensitivity $\delta r$ in the measurement of 
critical field must be of order $ \delta r = 
\frac{\Delta H_{||}}{H_{||}} \approx 5-10/1000$ in the temperature range of 
interest, and the sensitivity on measurement of the reduced 
temperature $\delta t$ of order $\delta t \leq 
\delta r(\frac{1}{H_{||}}\frac{\partial H_{||}}{\partial t})^{-1} 
\approx 3*10^{-4}$, corresponding to the sensitivity in absolute 
temperature $\delta T \approx 0.25$ mK. It is important to point out that, 
in this experiment, alignment requirements are quite stringent: from 
the formula  \cite{Tinkam} $\frac{\delta H_{||}}{H_{||}} = 
\frac{1}{H_{||}}\frac{\partial H_{||}}{\partial \theta} \delta \theta =
\frac{H_{||}}{H_{\bot}}\delta \theta$, were $\theta$ is the angle between 
field and sample plane, the penetration depth $\lambda$ is typically of 
order 50-100 nm. On considering a film thickness of 5 nm we obtain, near 
transition, $\frac{H_{||}}{H_{\bot}} \approx \sqrt{24}
\frac{\lambda}{D}\frac{1}{\sqrt{1 - t}}$ which imposes the stringent limit 
$\delta \theta < 3*10^{-5} $rad (Zinc/Gold), relaxed to 
$\delta \theta < 10^{-4}$ rad for Zinc and perfect mirror.

\section{Apparatus description and sensitivity tests}

The cryogenic apparatus consists of the commercial cryostat Oxford 
Instruments HELVLTD HelioxVL 3He inserted in a HD120H transport dewar, 
reaching the base temperature of 300 mK. The external field is generated by 
a 1.1 Gauss/mAmpere superconducting coil and the current is supplied and 
measured with a sensitivity better than 1/1000 by a multimeter HP 34401A. The 
sample can be oriented parallel (and orthogonal) to the magnetic field, 
aligned by construction with an estimated accuracy of about $10^{-2}$ rad.
Possible alignment improvements  will be discussed in the next section. \\
The measurement method is a standard omodine four-wire resistance. To test 
the sensitivity of cryogenic apparatus a film of 300 nm thickness has been 
used, so as to have far less stringent limitations on misalignments. 
The lock-in frequency is 6 Hz and probe current of $10 \mu A$.  The 
resistance of the film before transition is $R = 24 m\Omega$. The actual 
measurement is performed by fixing the external field and storing 
$R(T)$. A set of measurements is reported in fig. 3: the transition width 
is approximately 10 mK, the measured residual resistance is about 1 
$m\Omega$. The sensitivity of the measurement $\delta R \approx 1 m\Omega$ 
is limited by the noise current at first stage of the read-out electronic. 
The present limit is thought to be sufficient also for the final 
measurement, where thinner films will have about two orders of magnitude 
higher resistances. \\

\begin{figure}
	\centering
		\includegraphics[width= 8 cm]{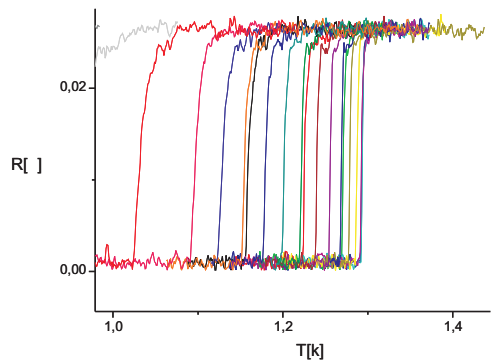}
	\caption{R(T) for diffrent applied field on a 300 nm Aluminum sample}
	\label{fig:transizioni}
\end{figure}

The experimental data, reported in fig. 4 in the temperature region of 
interest, show the expected behaviour $H_{||}(t) \propto \sqrt{1 - t}$. In 
order to estimate sensitivity in $\frac{\delta H_{||}}{H_{||}}$ the data 
have been fitted by taking into account that the correction resulting from
nucleation is not negligible, by virtue of sample thickness. Thus, the 
data have been fitted with
\begin{equation}
H_{||}(t) = \sqrt{24}H_{T}(0)\frac{\lambda_{e}(0)}{D}\sqrt{\frac{1 - t^{2}}{1 + t^{2}}}\left[1 +
\frac{9}{\pi^{6}}\frac{D^{2}}{{\overline \xi}(0)^{2}}(1-t)\right],
\end{equation}
which is valid near $T_{c}$, where the conditions $D < \sqrt{5} \lambda_{e}(t)$ 
and $\frac{9}{\pi^{6}}\frac{D^{2}}{{\overline \xi}(0)^{2}}$ are satisfied. 
In the equation $H_{T}(0)$, $\lambda_{e}(0)$ and $ \bar{\xi}(0)$ are 
the thermodynamical field, the effective penetration depth 
and the coherence length at zero temperature, respectively \cite{Tinkam}. \\

\begin{figure}
	\centering
		\includegraphics[width= 8 cm]{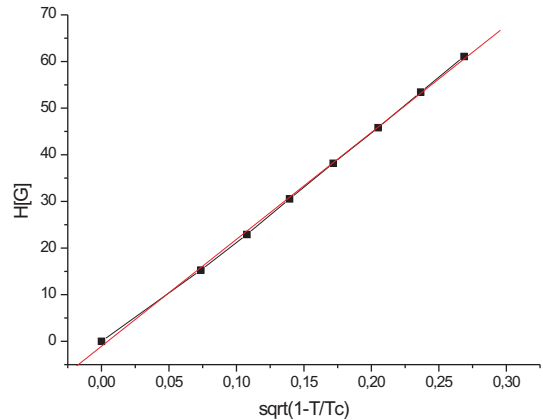}
	\caption{Critical field as a function of $\sqrt{1 -t}$}
	\label{fig:lineare}
\end{figure}

The results of the fit are shown in fig. 4, where the experimental residuals 
are reported. In the same figure the red curves are the confidence bands, 
the green curve is the expected signal for an Aluminum/Gold cavity, the 
purple for a Zinc/Gold cavity and finally the blue is the Zinc expected 
signal if the contribution of the zero-frequency Transverse Electric mode 
is set to zero in the normal state.\\
The residuals show a sensitivity $\frac{\delta H_{||}}{H_{||}} 
\approx 3*10^{-3}$ in the region of interest, so that $\delta t $ can be 
estimated as $\delta t \approx 1.5*10^{-4}$, corresponding to $\delta T 
\approx 0.2$ mK. Last, from the fit we obtain the values 
$T_{c} = 1.2932$ K, $\sigma_{T_{c}}= 0.0002$ K; $\lambda_{e}= 104.3$ nm,  
$\sigma_{\lambda_{e}}= 0.3$ nm;  
$ \overline{\xi}(0) = 60$ nm, $\sigma_{\overline{\xi}(0)}= 20$ nm K 
which are in the range of values compatible with the literature. \\

\begin{figure}
	\centering
		\includegraphics[width= 8 cm]{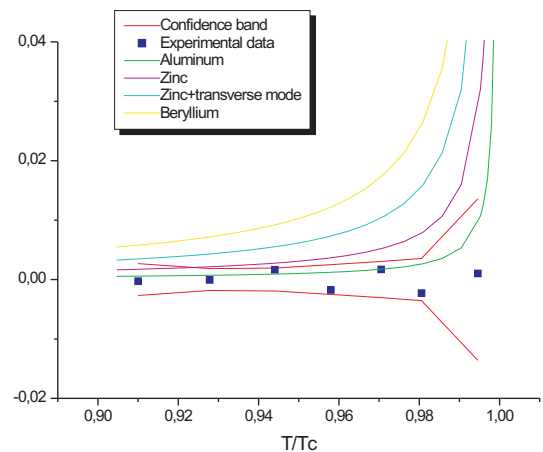}
	\caption{Fit Residuals}
	\label{fig:residui}
\end{figure}

From this measurement we find that the sensitivity of the measurement might 
not be sufficient to detect the Aluminum/Gold cavity signal; on the 
contrary, it should be sufficient to detect it in Zinc/Gold cavities, 
allowing also to discriminate the questioned contribution of zero frequency 
TE mode to Casimir energy \cite{Sernelius} \cite{Bimon2}.

\section{Next sensitivity test and experiment schedule}
  
Although these preliminary results on cryogenic apparatus are encouraging, 
various improvements should be performed before we can state that the needed 
sensitivity has been reached. In particular, two very important effects 
might spoil the present sensitivity: the broadening of transition width  
and rising of alignment effects when passing to thinner films. In this 
spirit, while first experimental studies on Zinc deposition and cavity 
realization are carried out, a first experimental test with aluminum 
cavities will be performed in a short time (fall of 2005). The thickness of 
Aluminum layer will be 10 nm, while the third layer will be Gold (100 nm) 
for some cavities and Silver (100 nm) for others. Gold is chosen for 
its extreme simplicity in deposition, while Silver, more reflective, will 
be tested as a candidate for final configuration material.\\
As discussed previously, the requirements on alignment are quite stringent. 
A solution that we will test in the next run is the use, on the same sample, 
of two simple films, two cavities and a bridge configuration, as shown in 
fig. 6.

\begin{figure}
	\centering
		\includegraphics[width= 8 cm]{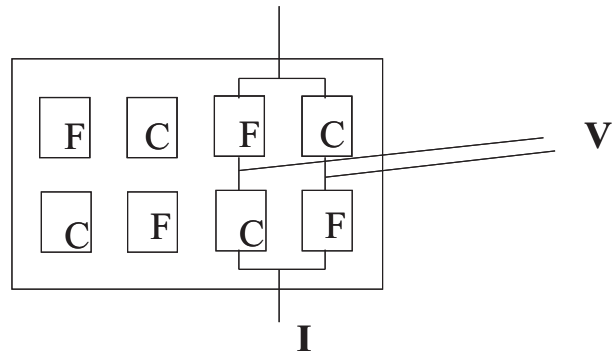}
	\caption{Scheme of the sample: statistics will be performed on different samples}
	\label{fig:ponte}
\end{figure}

 It is important to stress that our experiment looks for a different 
behaviour of $H_{||}(T)$ for films being or not being part of a cavity. 
Thus, this configuration is not aimed at improving accuracy on the 
measurement but rather at obtaining ``on line'' the different behaviour of 
the two cases which will have the same misalignment by construction. In 
this respect, we point out that we do not expect 
that the bridge will be compensated during the transition; 
it should instead exhibit a pick by virtue of imperfect equalities of the 
four samples: the evolution of the pick for different external applied 
field will be the desired evidence of the different behaviour of the 
film/cavity $R(T)$.\\
These structures are presently under construction at the IPHT (Jena) and 
the first measurements are foreseen by the end of the year 2005.      

\section*{Acknowledgments}

G Bimonte and G Esposito acknowledge partial financial support by 
PRIN {\it SINTESI}.

\end{document}